# Bed-inventory Overturn Mechanism for Pant-leg Circulating Fluidized Bed Boilers

Zhe Wang[1], Jining Sun[2], Zhiwei Yang[1], Logan West[1], Zheng Li[1*]

[1] State Key Lab of Power Systems, Department of Thermal Engineering, Tsinghua University, Beijing, 100084, China

[2] National Key Laboratory of Science and Technology on aero Engine Aero-thermodynamics, School of Jet Propulsion, BUAA, Beijing, 100191, China

## Abstract

A numerical model was established to investigate the lateral mass transfer as well as the mechanism of bed-inventory overturn inside a pant-leg circulating fluidized bed (CFB), which are of great importance to maintain safe and efficient operation of the CFB. Results show that the special flow structure in which the solid particle volume fraction along the central line of the pant-leg CFB is relative high enlarges the lateral mass transfer rate and make it more possible for bed inventory overturn. Although the lateral pressure difference generated from lateral mass transfer inhibits continuing lateral mass transfer, providing the pant-leg CFB with self-balancing ability to some extent, the primary flow rate change due to the outlet pressure change often disable the self-balancing ability by continually enhancing the flow rate difference. As the flow rate of the primary air fan is more sensitive to its outlet pressure, it is easier to lead to bed inventory overturn. While when the solid particle is easier to change its flow patter to follow the surrounding air flow, the self-balancing ability is more active.

## 1. Introduction

Circulating fluidized bed (CFB) boiler technology has been widely applied in China's power industry due to its intrinsic advantages, such as higher efficiency, lower pollution, and greater fuel flexibility, in comparison to pulverized coal boilers. In a CFB, the solid particles are lifted out of the bed or riser by the supplied air inflow, separated from the fluid by cyclone, and recycled to the bed. At present, the scale of CFB boilers is quickly accelerating for application in larger power plants. Whereas conventional CFB boilers are typically designed as straight shaft structures, as the scale of CFB boilers increases, they are more commonly designed with a pant-leg structure. For example, China has imported an Alstom designed, 300MWe pant-leg CFB boiler to the Baima power plant in Sichuan province and is now in the process of designing and manufacturing a 600MWe pant-leg CFB boiler as a demonstration project.

The pant-leg design is typical for higher capacity boilers because it allows for better secondary air penetration, maintaining good air-coal mixing as well as efficient combustion. The pant-leg structured CFB boiler features two separated legs with solid particles fluidized by independent primary air supplies. This newly-designed feature leads to unique CFB dynamic performance. One special occurrence is the lateral transfer of bed-inventory between the two legs, a critical process tightly coupled with combustion, heat transfer, and flow conditions. Lateral bed-inventory transfer greatly affects the dynamic performance of the CFB and can potentially

---

lead to unexpected accident, bed-inventory overturn, during operation.

To understand the mechanisms underlying the lateral mass transfer and to avoid bed-inventory overturn, Li et al.[1,2] experimentally studied the lateral transfer of solid particles in a bench-scale, cold CFB riser with pant-leg structure. A compounded pressure drop mathematic model was developed, and it was concluded that the main reason for lateral transfer of solid particles is the lateral pressure gradient of the gas phase in the CFB and once the pressure balance is broken, it rarely returns back to balance without auto-controlling the primary air fan. Until now, little additional research has addressed the working mechanism of bed-inventory overturn and detailed heterogeneous flow behavior such as lateral mass transfer.

The flow, combustion, and heat transfer processes inside the CFB are very complicated. Earlier one-dimensional (1-D) simulation models, such as Li and Kwauk[3], Kunii and Levenspiel[4], and Smolders and Baeyens[5] only compute the axial variation of solid holdup, combustion, and heat transfer, while neglecting the radial variations. These models are able to predict the axial behavior of solid density in the riser without considering radial distribution.

More comprehensive models were developed to describe the radial variation of the density of solids, by solving the conservation equations of mass and momentum, namely the Navier-Stokes equations, for each phase[6-8]. In the generally applied Eulerian-Eulerian approach, the Newtonian rheological model is extensively applied[9]. A key step of such an approach is to determine the thermodynamic and transport properties of the solid phase. One procedure in determining these properties makes use of empirical relations[10-13] from experimental data. With such models, Gungor and Eskin[14] predicted the occurrence of core-annular flow in a axial symmetric CFB riser and the axial and radial distribution of solid volume fraction, void fraction and particle velocity were in agreement with atmospheric cold bed CFB units' experimental data given in the literature. Another approach applied recently for determining the thermodynamic and transport properties is the kinetic theory of granular flow (KTGF). Almuttahar and Taghipour[15] evaluated the application of KTGF comprehensively in CFB modeling by comparing predictions of a CFB riser at various fluidization conditions with experimental results and concluded that the calculated solid volume fraction and axial particle velocity were in good agreement with the experimental data within a high density, fast fluidization regime. Lu et. al.[16] modeled a three-phase flow of gas, solid particles, and clusters in a CFB riser. The computed solid mass fluxes and volume fractions agreed well with their experimental data. Many other authors, such as Benyahia et al.[17], Neri and Gidaspow[18] and Zheng et al.[19] have also achieved reasonable predictions of gas-solid hydrodynamics using KTGF. The KTGF model has become one of the most useful tools for modeling fluid-solid flows in dilute to dense bed regimes.

Therefore, in the present work, the Eulerian-Eulerian approach with KTGF is applied to model the flow characteristics of a pant-leg structured CFB boiler. The aim of this work is to model the detailed hydrodynamics of a pant-leg CFB boiler to deeply understand the flow characters inside the pant-leg CFB as well as the bed-inventory overturn mechanism.

## 2. Model description

While recognizing that such complex systems inside CFB risers are better studied in three dimensions (3-D) to capture the detailed picture of the flow, given the current computational power, two-dimensional (2-D) models remain more popular[20,21]. Since the flow behaviors along the depth of the riser are similar, for this paper, a 2-D model instead of a 3-D model was designed to minimize the computing time while providing enough information for understanding the flow patterns.

For this model, the 300MWe pant-leg CFB boiler designed by Alstom at Baima power plant was chosen as the prototype. The structure and dimensions of the pant-leg CFB boiler 2-D model are shown in Fig. 1.

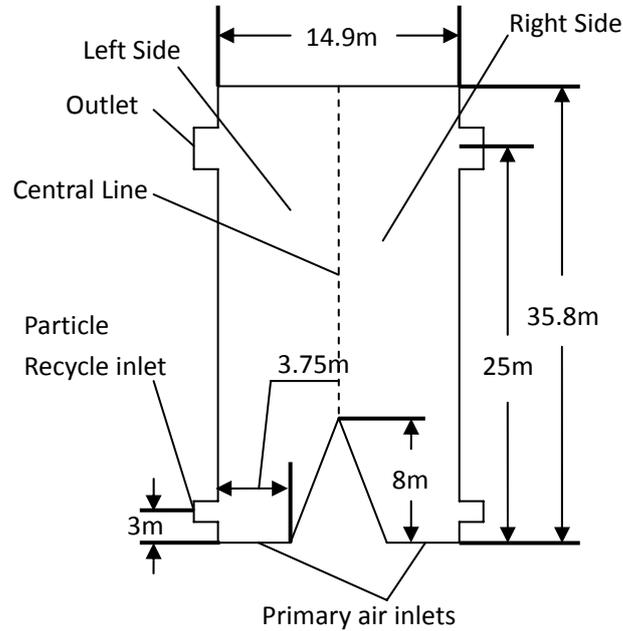

*Figure 1: Diagram of pant-leg CFB boiler*

In real CFB operation, the primary air is boosted into the riser at the bottom of each leg to fluidize the solid particles inside the boiler. As the solid materials are fluidized, part of them are carried out of the riser, separated from the air flow by cyclone, and returned to the boiler via recycle ducts. By this mechanism, bed-inventory in CFB boilers is conserved. In the present work, the detailed flow pattern inside the cyclone and the recycle ducts was not modeled. Instead, a user defined function (UDF) was applied to simply separate the solid particles from the gas flow at the outlet and send them back to the riser through the particle recycling inlets. The UDF calculates the volume fraction of solid particles at the solid particle recycle inlets by gathering the solid particles out of the riser at the correspondent outlet. The velocity of solid particles was set to be as low as 0.1 m/s to reduce its impact on the hydrodynamics behavior in the CFB riser. The outlets boundary was set to be a pressure outlet with constant atmospheric pressure. In addition, the effects of secondary air flow were neglected in the present work, although secondary airflow could lead to more or less lateral mass transfer and therefore affect the bed-inventory overturn process in real application.

*2.1 Governing equations*

An Eulerian-Eulerian two-fluid model with KTGF was used to simulate the detailed hydrodynamic behavior of the flow in this 2-D pant-leg structured CFB riser. One set of the mass and momentum conservation equations were solved for each phase, where the momentum equations were linked by an inter-phase exchange term. In the case of the isothermal condition without mass transfer, as considered in this study, the governing equations for the gas and solid phase are as follows[22].

The continuity equations for the gas and solid phase, g and s, respectively, are described by

$$\frac{\partial}{\partial t}(\varepsilon_g \rho_g) + \nabla \cdot (\varepsilon_g \rho_g \vec{u}_g) = 0 \tag{1}$$

and

$$\frac{\partial}{\partial t}(\varepsilon_s \rho_s) + \nabla \cdot (\varepsilon_s \rho_s \vec{u}_s) = 0 \tag{2}$$

where $\varepsilon$, $\rho$ and $u$ are the volumetric fraction, density and velocity of each phase, respectively, and the subscripts $g$ and $s$ are gas and solid.

The momentum equation for the gas phase is given by

$$\frac{\partial}{\partial t}\left(\varepsilon_g \rho_g \vec{u}_g\right) + \nabla \cdot \left(\varepsilon_g \rho_g \vec{u}_g \vec{u}_g\right) = -\varepsilon_g \nabla p + \nabla \cdot \overline{\overline{\tau}}_g + \varepsilon_g \rho_g \vec{g} + \beta(\vec{u}_s - \vec{u}_g) \tag{3}$$

and for the solid phase is given by

$$\frac{\partial}{\partial t}\left(\varepsilon_s \rho_s \vec{u}_s\right) + \nabla \cdot \left(\varepsilon_s \rho_s \vec{u}_s \vec{u}_s\right) = -\varepsilon_s \nabla p + \nabla \cdot \overline{\overline{\tau}}_s + \varepsilon_s \rho_s \vec{g} + \beta(\vec{u}_g - \vec{u}_s) \tag{4}$$

where $p$ is the fluid pressure, $\tau$ is the stress tensor, $g$ is gravity, and $\beta$ is the gas-solid drag coefficient.

The conservation equation for the fluctuation energy of the solid phase, known as granular temperature, $\Theta$, can be obtained by solving its transport equation:

$$\frac{3}{2}\left[\frac{\partial}{\partial t}\left(\varepsilon_s \rho_s \Theta\right) + \nabla \cdot \left(\varepsilon_s \rho_s \vec{u}_s \Theta\right)\right] = \overline{\overline{\tau}}_s : \nabla \vec{u}_s - \nabla \cdot (k_\Theta \nabla \Theta) - \gamma_\Theta + \Phi_\Theta \tag{5}$$

where $k_\Theta$ is the thermal diffusion coefficient, $\gamma_\Theta$ is the collision dissipation energy, and $\Phi_\Theta$ is the transfer of kinetic energy between gas and solid phases. Definitions of these terms are provided in Table 1.

**Table 1** Constitutive equations for the fluctuation energy of solid phase

$$k_\Theta = \frac{150 \rho_s d_p \sqrt{\Theta \pi}}{384(1+e)g_0}\left[1 + \frac{6}{5}(1+e)\varepsilon_s g_0\right]^2 + 2\rho_s \varepsilon_s^2 d_p (1+e)g_0 \sqrt{\frac{\Theta}{\pi}} \tag{6}$$

$$\gamma_\Theta = \frac{12(1-e^2)g_0}{d_p \sqrt{\pi}} \rho_s \varepsilon_s^2 \Theta^{3/2} \tag{7}$$

$$\Phi_\Theta = -3\beta\Theta \tag{8}$$

The stress tensor for the gas phase, $\bar{\bar{\tau}}_g$ can be expressed as

$$\bar{\bar{\tau}}_g = \varepsilon_g \mu_g \left(\nabla \vec{u}_g + \nabla^T \vec{u}_g\right) + \varepsilon_g \left(\lambda_g - \tfrac{2}{3}\mu_g\right) \nabla \cdot \vec{u}_g \bar{\bar{I}} \tag{9}$$

For the solid phase, $\bar{\bar{\tau}}_s$ can be described by

$$\bar{\bar{\tau}}_s = -p_s \bar{\bar{I}} + \varepsilon_s \mu_s \left(\nabla \vec{u}_s + \nabla^T \vec{u}_s\right) + \varepsilon_s \left(\lambda_s - \tfrac{2}{3}\mu_s\right) \nabla \cdot \vec{u}_s \bar{\bar{I}} \tag{10}$$

where $\mu$, $\lambda$ and $I$ are shear and bulk viscosities and unit tensor, respectively. Definitions of $\mu_s$ and $\lambda_s$ are given in Table 2.

**Table 2** Constitutive equations for the momentum of solid phase

$$\mu_s = \frac{4}{5}\varepsilon_s \rho_s d_p g_0 (1+e)\sqrt{\frac{\Theta}{\pi}} + \frac{10\rho_s d_p \sqrt{\Theta \pi}}{96\varepsilon_s (1+e) g_0}\left[1 + \frac{4}{5}(1+e)\varepsilon_s g_0\right]^2 \tag{11}$$

$$\lambda_s = \frac{4}{3}\varepsilon_s \rho_s d_p g_0 (1+e)\sqrt{\frac{\Theta}{\pi}} \tag{12}$$

The solid phase pressure, $p_s$, is described in the context of the KTGF as:

$$p_s = \varepsilon_s \rho_s \Theta + 2(1+e)\rho_s \varepsilon_s^2 g_0 \Theta \tag{13}$$

where $e$ is the particle-particle restitution coefficient and $g_0$ is the radial distribution function, as given by

$$g_0 = \left[1 - \left(\frac{\varepsilon_s}{\varepsilon_{sm}}\right)^{1/3}\right]^{-1} \tag{14}$$

where the subscript $m$ is maximum.

The gas-solid drag inter-phase exchange coefficient is expressed as

$$\beta = \begin{cases} \dfrac{150\varepsilon_s^2 \mu_g}{\varepsilon_g d_p^2} + \dfrac{1.75\rho_g \varepsilon_s |\vec{u}_g - \vec{u}_s|}{d_p} & \varepsilon_g \leq 0.80 \\ \dfrac{3C_D \varepsilon_g \varepsilon_s \rho_g |\vec{u}_g - \vec{u}_s|}{4d_p} \cdot \varepsilon_s^{-2.65} & \varepsilon_g > 0.80 \end{cases} \tag{15}$$

where $d_p$ is the diameter of the solid particles and $C_D$, the drag coefficient, is given by

$$C_D = \begin{cases} \dfrac{24\left(1+0.15(\mathrm{Re}_s)^{0.687}\right)}{\mathrm{Re}_s} & \mathrm{Re}_s < 1000 \\ 0.44 & \mathrm{Re}_s \geq 1000 \end{cases} \quad . \tag{16}$$

The solid Reynolds number, $\mathrm{Re}_s$, is expressed as

$$\mathrm{Re}_s = \frac{\rho_g \varepsilon_g \left|\vec{u_g} - \vec{u_s}\right| d_p}{\mu_g} \quad . \tag{17}$$

The commercial CFD package (Fluent Inc., V12.0) was used to provide a numerical solution for the governing equations. The finite volume method[23] was applied to discretize the governing equations. A second-order upwind discretization scheme was used to solve the convection terms. A convergence criterion of $1\times10^{-5}$ was specified for the relative error of the successive iterations.

*2.2 Simulation Conditions*

In order to best isolate and analyze the controlling factors for CFB hydrodynamics and bed-inventory overturn, this work studied the flow scenarios under different conditions. Starting from the same initial conditions, the flow behavior of the fluid should be completely determined by the boundary conditions as well as the gas and solid properties. For pant-leg CFB application, the most important boundary condition is the primary air flow rate, and the most important fluid and solid property is the gas-solid interaction. Therefore, thorough work was conducted to investigate the impact of the primary air flow and gas-solid interaction.

With respect to the primary air flow, the present work considered the conditions where there was an independent primary air fan for each leg of the pant-leg CFB with identical performance curves. In total, four different types of primary air fans were investigated. Fan A represents ideal fans that are able to provide constant air flow even with varying outlet pressure. Fan B stands for air fans with performance curves such that, as the outlet pressure increases, the air flow rate of the fan decreases. Fan C and D stand for more realistic Roots blowers that have performance curves for which flow rate does not change with outlet pressure change at low pressure. The difference between C and D is that D maintains a constant air flow velocity over a wider range of outlet pressures. The performance curves of the four primary air fans are shown in Fig. 2.

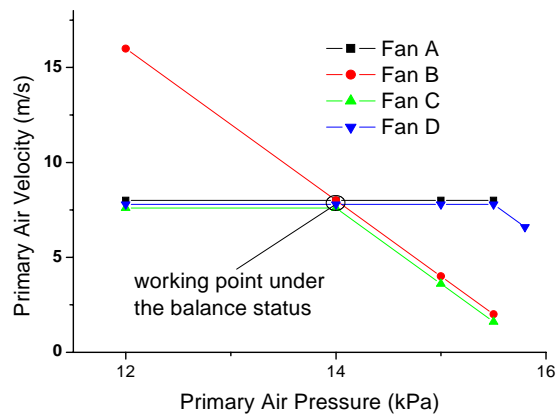

*Figure 2. Schematic working characteristic lines of the four primary air fans.*

With respect to the gas-solid interaction, a key parameter is the diameter of the solid particles. In this work, the diameter typically applied was 0.15 mm, but to show its impacts on CFB operation, diameters of 0.0725 mm and 0.05 mm were also studied.

For the simulation, the computational domain consisted of about 30,000 grids. As the grids increase to 40,000, the results (e.g. solid distribution) do not change significantly, indicating mesh independent results have reasonably been achieved with the mesh number equal to or greater than 30,000. The time step was $5\times10^{-3}$ s and the simulation time lasted for 200 s of real fluidization time, corresponding to 5-7 days of computing time on a 3 GHz workstation.

## 3. Results and discussion

### 3.1 Instability and self-balancing of bed-inventory in pant-leg CFB

Starting from the condition where all bed-inventory was symmetrically piled on the bottom of the pant-leg CFB, as shown in Fig. 3a, and applying constant and equal primary air flow, as using Fan, to fluidize the bed-inventory, the simulation showed that during the early stage before 10 seconds (Fig. 3b) the two-phase flow structure of the pant-leg CFB was symmetric. Then, three seconds later, it reached a break point where the flow structure became slightly asymmetric (Fig. 3c). From this point on, an asymmetric and unsteady flow pattern continuously existed (Fig. 3d, 3e, and 3f). While disturbance is inevitable for real CFB operation, this result shows that for the symmetrically designed pant-leg CFB, even under perfectly balanced air supply conditions, it is not possible to maintain an absolutely steady state operation; instability is inevitable. Therefore, a relatively steady state (Fig. 3e) was chosen as the baseline condition for later calculations simulating other fans unless explicitly stated otherwise. In addition, to obtain similar bed-inventory and pressure conditions with the real riser, iterations were made to produce a suitable pressure-drop (about 14 kPa) over the riser before taking this series of calculations for demonstration, by changing the height of the initial piled bed-inventory.

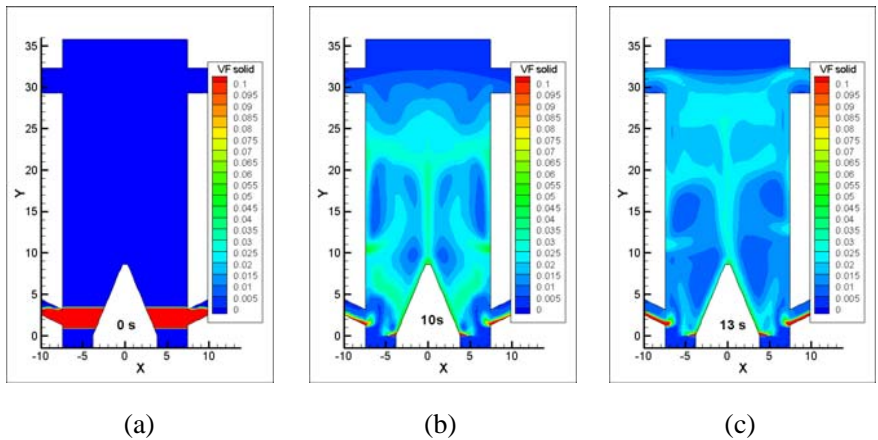

(a)  (b)  (c)

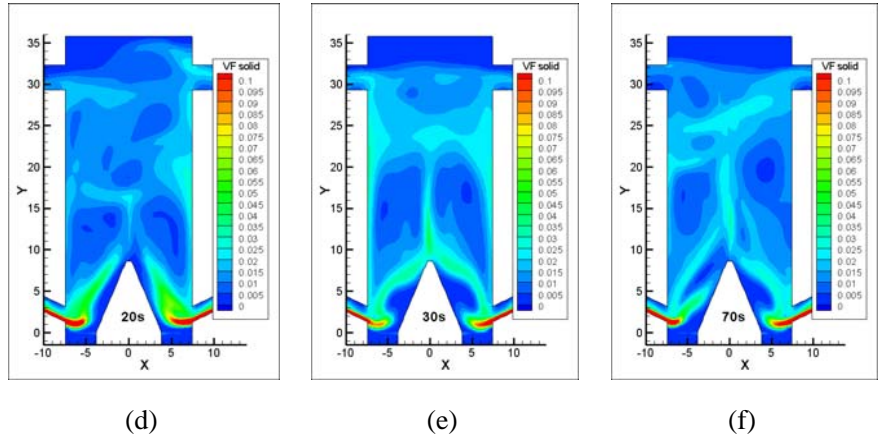

(d)                    (e)                    (f)

*Figure 3. Contour of the volume fraction of solid particles inside the CFB with different operational time (s) applying Fan A.*

Figure 4 shows the profile of the total pressure-drops over each side of the riser and the bed-inventories in each side of the riser varying over time during operation. Both the total pressure-drop of each side of the riser and the bed-inventories in each side of the riser fluctuated around the average values. In addition, it has been noticed that what directly affect lateral bed-inventory transfer is the pressures difference in the upper regions of the boiler instead of the bottom pressure difference. But, since there is strong correlation between the upper location and bottom pressures as well as pressure difference, in calculating the pressure swings, we applied the pressure the bottom of the CFB leg around the air fan outlet as the representative.

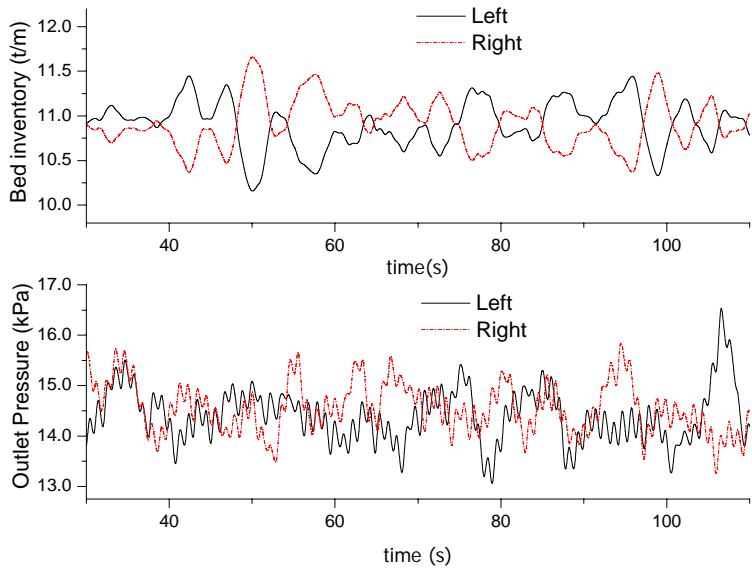

*Figure 4. Profile of the total pressure-drops over the riser and the bed inventories in the riser varying with operational time while applying Fan A.*

The reason behind the instability and fluctuations can be explained by analyzing the situation illustrated in Fig. 3. Since the primary flow rate was constant with Fan A, the outlet pressure of the primary air fan was influenced mainly by flow resistance. Flow resistance, in turn, was determined by the profiles of the volume fraction and the velocity of the solid particles. Flow resistance was higher in the regions in which the volume fraction of solid particles was higher or in which the solid particles possessed velocities opposite to air flow. The profile of the volume fraction of the solid particles was unsteady as well as asymmetric (shown in Fig. 3), and the velocities of the solid particles were in a similar state, too. Therefore, the flow resistance on the left side and the right side of the riser always varied, causing the fluctuation of the respective outlet pressures.

Furthermore, the flow structure in the pant-leg structured CFB was a typical "core annular" structure in which the solid particles flowed upward near the central line and downward near the two side walls. The distinguishing feature between pant-leg structured CFBs and conventional CFBs was that, in pant-leg structures, the volume fraction of solid particles near the central line was higher (Fig. 3) and the solid particles swing laterally, leading bed-inventory in each side to vacillate greatly. This concentration of solid particles near the central line was important to pant-leg CFBs as it influenced the lateral transfer rate of bed-inventory and hence the pressure-drop of each side of the riser.

Generally, larger bed-inventory inside the CFB boiler leads to higher flow resistance and outlet pressures of the primary air fans, but the distribution and velocities of the solid particles also exert important influence. Under conditions in which solid particles move from the right to the left side, the bed-inventory and the pressure-drop of the left side of riser increase while the pressure-drop of the right side decreases. The resulting pressure difference between the two sides inhibits the continue movement of solid particles. However, due to particle inertia, the lateral movement of the solid particles does instantly not stop upon encountering opposing forces; some solid particles continue their movement. With two identical A type primary air fans, the pressure difference increases until lateral transfer of the solid particles is entirely prevented by the pressure difference. According to Newton's second law, the time difference between the onset of the pressure difference and the actual prevention of solid particle motion is determined by the gas-solid interaction and the inertia of solid particles (which would be discussed further in section 3.3). After solid particle transfer to the left side ceases entirely, the pressure difference drives the solid particles back to the right side. This lateral transfer process then fluctuates from side to side, causing the vacillation of the total pressure-drop and bed inventories of each side.

The process described above typifies a negative feedback system in which buildup of bed-inventory on one side due to lateral mass generates pressure gradient that reverses the trend of lateral mass transfer. The intrinsic instability is attributable to the high volume fraction of solid particles near the central line and the time difference between the onset of the pressure difference and the actual prevention of solid particle motion. The typical response time for negative feedback to generate reversal is determined by the gas-solid interaction and the inertia of solid particles. Because this slowly built-up negative feedback process enables the CFB boiler to consistently return to a balanced state, these results indicate that pant-leg CFB boilers can be considered as having a self-balancing nature.

*3.2 The impact of primary air fan performance curves*

As the primary air fan is a critical boundary condition controlling flow in pant-leg structured CFB, the fan properties greatly influence the dynamic features of pant-leg CFB. Three different types of air fans (Fan B, C, D as in Fig. 2) were investigated, applying each kind of fan to the initial conditions present in case with Fan A (Fig. 3e).

When Fan B was applied to the two primary air inlets, the bed-inventory of left side of the riser continually increased from 40s while the right side continually decreased (Fig. 5). In this case, due to the performance curve of fan B, flow velocity from the left primary air fan kept decreasing in response to the higher outlet pressure as bed-inventory moved in. At its upper limit (16kPa), flow velocity dropped to zero. At the same time, the working point of the right side primary air fan moved to the lower pressure region, causing its flow velocity to increase (Fig. 2). The subsequent imbalance in flow velocities brought more solid particles from the right side to the left side until complete overturn occurred. The process of bed-inventory overturn can be seen in the flow structures shown in Fig. 6.

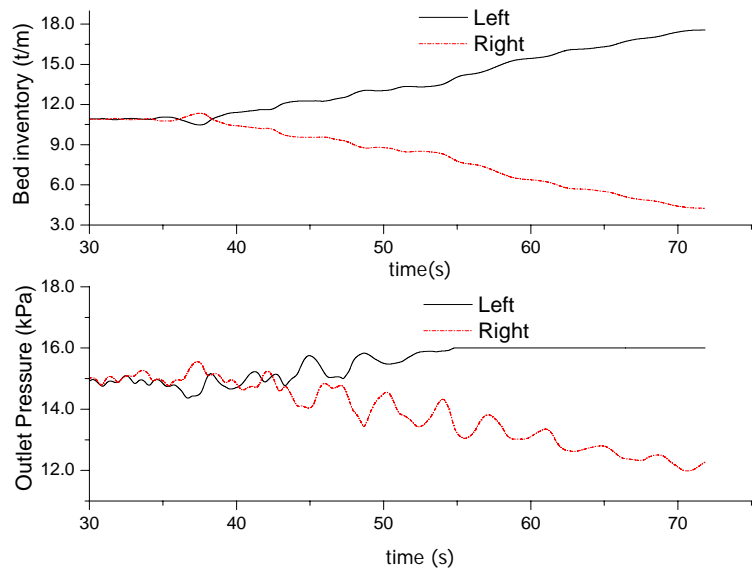

Figure 5. *The profile of the total pressure-drop over the riser and the bed-inventory in the riser varying over operational time applying Fan B.*

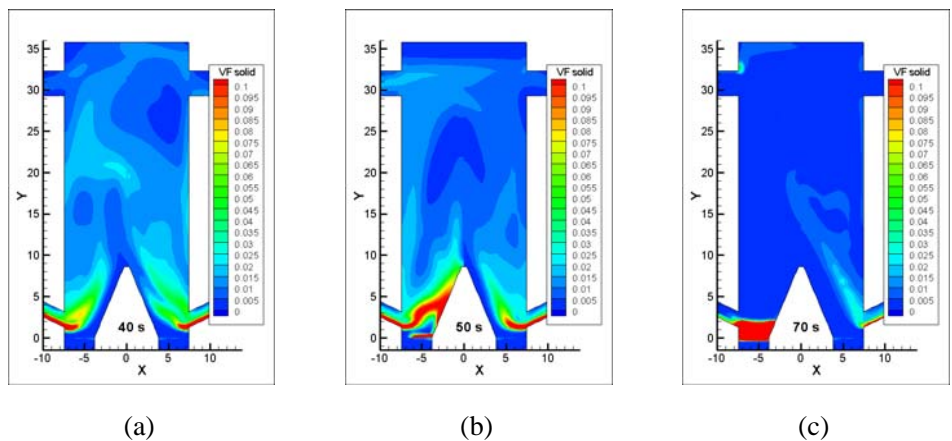

(a)  (b)  (c)

Figure 6. *Contour of the volume fraction of solid particles inside the CFB with different operational time (s) applying Fan B.*

The response of the primary air fan to the outlet pressure change, which mainly comes from the increasing bed-inventory due to lateral mass transfer, serves as a positive feedback for mass transfer. That is, the primary air flow rate difference and the lateral transfer of solid particles promote each other. In pant-leg CFB boilers, this positive feedback process always works together with the negative feedback process described in section 3.1. Since it takes time to build-up enough lateral pressure difference to stop the mass transfer, the instantly change of the flow rate, as with fan B, supported the continuation of the lateral mass transfer. That is, the positive feedback develops faster than the negative feedback process, making the solid particles transfer from one side to another continually until bed-inventory overturn.

In addition, in real pant-leg CFB applications, there may be only one Roots blower to provide primary air for both legs from a conjunctional wind box as in Baima CFB power plant. Under this condition, as outlet pressure of one leg increase, the flow rate of the leg decrease and that of another leg decreases simultaneously since the total flow rate keeps constant. This is similar to the case with Fan B. Meanwhile, the simulation result of the case with Fan B has the same profile as the real operation when the overturn accident occurred[24]. Seen from the Fig. 5 and 6, bed-inventory overturn is inevitable and requires auto-control system to control the primary flow rate at all times, a finding which is in agreement with previous research [1].

While fans A and B are two extreme cases for identifying the mechanisms of bed-inventory overturn, simulations with Fans C & D represent more realistic scenarios in illustrating the CFB flow structure and mechanisms under typical fan conditions. The difference between Fan C and D is that Fan C operates at a constant velocity under a narrower pressure range than fan D (Fig. 2).

Figure 7 shows that bed-inventory fluctuated over a greater range with Fan C than Fan D. Furthermore, with Fan C, the average bed-inventory over the period from 30 to 70s was notably less on the left than the right side. This indicates that some amount of bed-inventory was permanently transferred to the left side, especially after 50s. Thus, as time progressed, there was a tendency towards bed-inventory overturn. For Fan D, however, this tendency did not appear over the time interval simulated, and the average bed-inventory was close to equal.

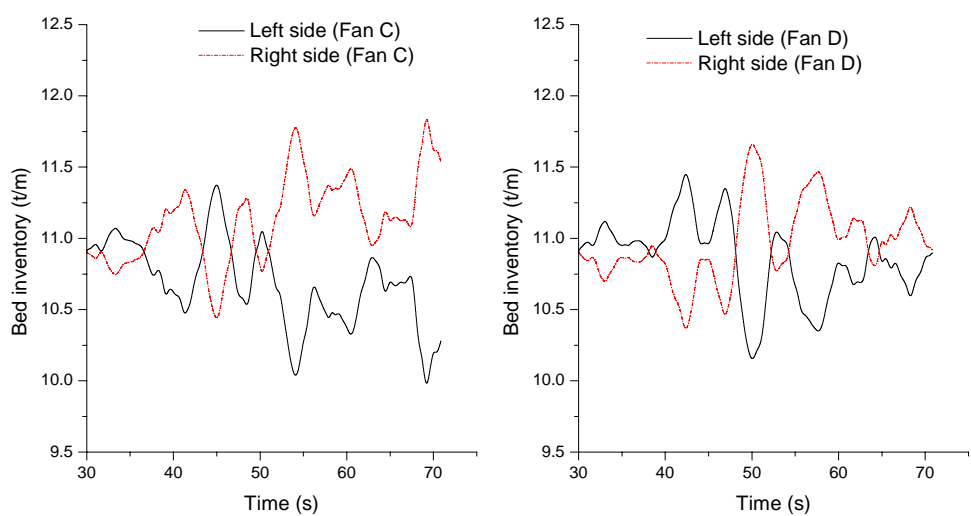

*Figure 7. The profile of the bed-inventory in the riser varying over operational time applying Fan C and D.*

The case with Fan D was similar to Fan A, since the outlet air pressure for the simulation with Fan D never reached the point above which the flow rate of the fan decreases. Therefore, the negative feedback process maintained the system self-balanced. For Fan C, there was a tendency towards bed-inventory overturn, indicating that the positive feedback process start to take over, as shown in Fig. 7.

The results demonstrate that the performance curves of the fans have an important influence on the process of lateral transfer in pant-leg structured CFB. The larger or faster the fan response to the outlet pressure change, the higher lateral transfer rate and the faster the bed-inventory overturn.

*3.3 The impact of gas-solid interaction*

As discussed in Section 3.1, a typical response time of the negative feedback process is determined by the gas-solid interaction and the inertia of solid particles. This delay time is determined by the property of the particles of how quickly to adjust themselves to follow the local air flow velocity. The ratio of drag force to particle inertial force characterizes the property. Under conditions that the particles have a larger ratio, there is less response time needed for the particles to adjust to and coincide with air flow, giving the fluid-solid two phase flow a more uniform flow structure inside the CFB boiler. Seen from Eq. 15, the diameter of solid particles is very critical to the ratio. The smaller the diameter of solid particles, the larger the ratio, the briefer the response time, and the more uniform the flow structure.

Based on the case with Fan B, two more simulations were conducted in which the diameters of the solid particles were varied. Figure 8 shows the profile of bed-inventory variation over time of simulation with respect to the different particle diameters. Compared to Fig. 5, the lateral transfer of solid particles was limited as the diameter decreases to 50%. Although the tendency of bed-inventory overturn still existed, the extent of bed-inventory overturn decreased. When the diameter was decreased to 30%, the lateral transfer of bed-inventory further decreased and the tendency of bed-inventory overturn was unnoticeable.

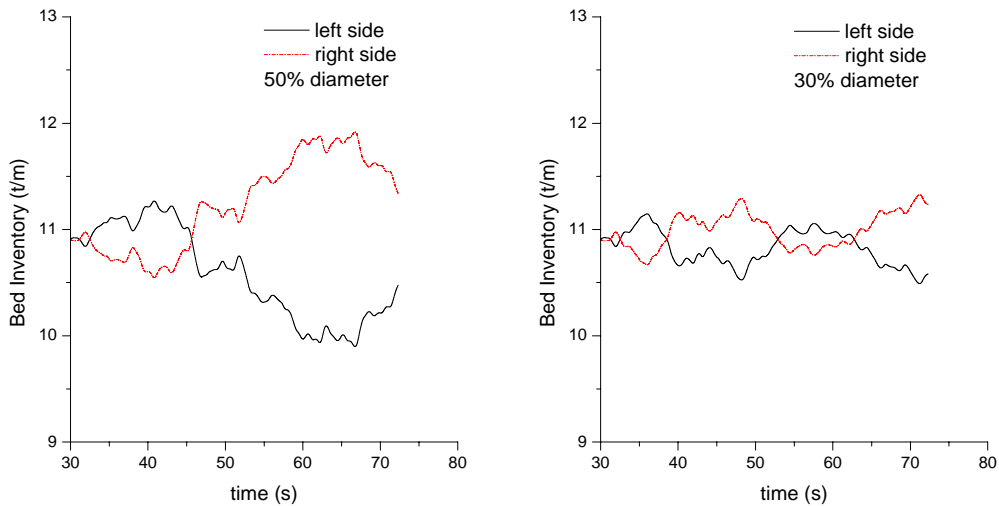

*Figure 8. The profile of the bed-inventory in the riser varying*
*with operational time applying Fan C and D.*

Figure 9 illustrates the volume fraction of 50% diameter solid particles inside the CFB at different operational times while applying Fan B. Compared Fig. 3 and 6, which are results for the original diameter, the volume fraction of solid particles is more uniform inside the CFB as analyzed above.

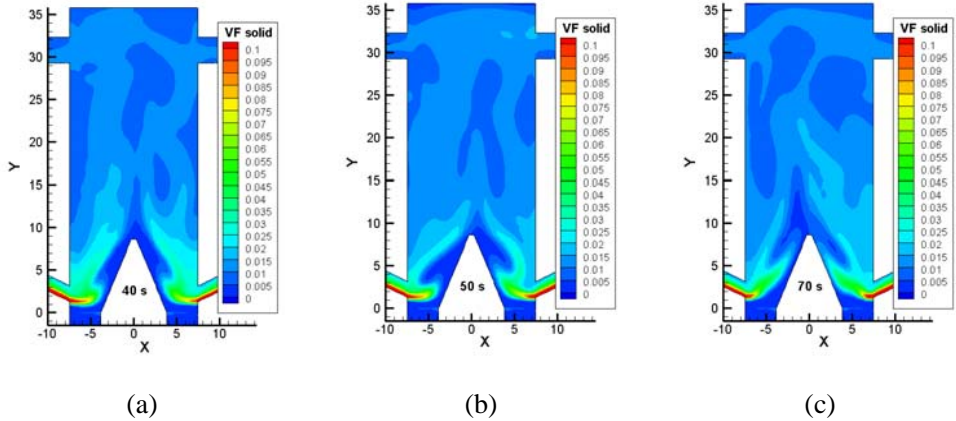

|    (a)    |    (b)    |    (c)    |

Figure 9. Contour of the volume fraction of 50% diameter solid particles inside the CFB
with different operational time (s) applying Fan B.

When the diameter decreases, the solid particles respond more quickly to its surround flow velocity, which is affected greatly and quickly by the lateral pressure difference. Therefore, the lateral bed-inventory transfer from one side to the other side stops faster and the fluctuating amplitude of bed-inventory and primary fan outlet pressure is reduced. This also decreases the impact of the positive feedback process since there is smaller outlet pressure change for the primary air fans. The combination of these effects greatly reduces the tendency for bed-inventory overturn.

**4. Conclusion**

A numerical model based on the Eulerian-Eulerian approach with KTGF was established to investigate the flow characteristics of a pant-leg structured CFB boiler as well as it bed-inventory overturn mechanism.

Similar to conventional CFB boilers, a core-annular flow structure exists inside the pant-leg CFB without ever exhibiting steady state flow. Unlike conventional CFB boilers, the volume fraction of solids along the central line of the pant-leg CFB is high, making it possible a large lateral mass transfer rate between the two legs and more possible of the occurrence of overturn.

The mechanisms for bed-inventory overturn are attributable to negative and positive feedback processes affecting the lateral bed-inventory transfer inside a pant-leg CFB boiler. The pressure difference caused by the lateral mass transfer inhibits the continuation of the movement, acting as a negative feedback process. At the same time, pressure differences on air flow outlets can cause differences in primary air flow rates and enhance the lateral mass transfer, acting as a positive feedback process. The negative feedback process, to some extent, provides the pant-leg CFB with self-balance ability, while the positive feedback process destroys this ability. These two processes work with each other and determine whether or not bed-inventory overturn occurs. Generally, if the air flow rate changes with the outlet pressure, the positive feedback process will develop faster than the negative one and most possibly lead to bed-inventory overturn. In the case that the negative process acts faster such as with very small particles, the CFB will have less possibility to lead to bed-inventory overturn.

The working mechanism of the primary air fans influenced bed-inventory overturn significantly. The performance curves of fans have an important influence on the positive feedback process of bed-inventory lateral transfer in pant-leg structured CFB. The larger the difference in air flow by each fan in response to change in the outlet pressure, the faster the positive feedback process expands, the higher the lateral transfer rate, and the faster the bed-inventory overturn. Fans that can maintain constant air flow over the widest range of outlet pressures were the ones least likely to overturn.

The property of how quickly the particles change its movement to respond to its surrounding air flow has important influence on both the negative feedback process and the positive feedback process. The reduction in the diameter of solid particles leads to a more uniform flow structure, smaller fluctuating amplitudes of bed-inventory, and ultimately lower possibility for bed-inventory overturn.

**Acknowledgement:** The work is sponsored as a national research project of China under contract 2005CB221207 and 2010CB227006.